\documentclass[aps,prl,showpacs,twocolumn]{revtex4}
\usepackage{graphicx}
\usepackage{amssymb}
\usepackage{amsmath}
\usepackage{color}

\def\bA{{\bf A}}
\def\bE{{\bf E}}

\def\br{{\bf r}}

\def\bxi{{\boldsymbol \xi}}
\def\bzeta{{\boldsymbol \zeta}}

\def\hb{\hat b}

\input{epsf}
\begin{document}

\title{Fermionic and bosonic ac conductivities at strong disorder}

\author{S.V.~Syzranov$^{1,2}$, O.M.~Yevtushenko$^{3}$, K.B.~Efetov$^{2}$}
\affiliation{$^1$Institute for Theoretical Condensed Matter Physics,
  Karlsruhe Institute of Technology,
  76131 Karlsruhe, Germany\\
  $^2$Theoretische Physik III, Ruhr-Universit\"at Bochum, 44780 Bochum, Germany\\
$^3$ Arnold Sommerfeld Center for Theoretical Physics and Center for Nano-Science,
Ludwig-Maximilians-University, 80333 Munich, Germany}
\date{\today}

\begin{abstract}
We study the ac conduction in a system of fermions or bosons
strongly localised in a disordered array of sites with short-range interactions at frequencies larger
than the intersite tunnelling but smaller than the characteristic fluctuation of the
on-site energy. While the main contribution $\sigma_0(\omega)$ to the conductivity comes from
local dipole-type excitations on close pairs of sites, coherent processes on three or more sites lead to an interference
correction $\sigma_1(\omega)$, which depends on the statistics of the charge carriers and
can be suppressed by magnetic field.
For bosons the correction is always positive, while for fermions
it can be positive or negative depending on whether the conduction is dominated by effective single-particle
or single-hole processes. 
We calculate the conductivity explicitly assuming a constant density of states of 
single-site excitations. Independently of the statistics, $\sigma_0(\omega)=const$.
For bosons $\sigma_1(\omega)\propto \log(C/\omega)$.
For fermions $\sigma_1(\omega)\propto\log[\max(A,\omega)/\omega]-\log[\max(B,\omega)/\omega]$,
where the first and the second term are respectively the particle and hole contributions,
$A$ and $B$ being the particle and hole energy cutoffs.
The ac magnetoresistance has the same sign as $\sigma_1(\omega)$.
\end{abstract}

\pacs{71.10.-w, 77.22.Gm, 74.25.N, 67.85.-d}


\maketitle

Interestingly, interference and many-body phenomena 
 in strongly disordered insulators
may be comparable to or even greater than those in clean materials.
%
Long ago fermionic many-body cotunnelling at strong disorder has been addressed
in the context of variable-range hopping in semiconductors\cite{ShklovskiiSpivak:review}.
The advent of modern cold-atom techniques\cite{Bloch:review}
and recent experiments on disordered superconductive films
\cite{Shahar:superinsulator,Baturina:superinsulator, Gantmakher:SITreview}
offered ample new opportunities to
observe interference and many-body effects at strong disorder, yet access them 
in both fermionic and bosonic systems.
Recently it was shown in Ref.~\cite{Syzranov:RG} and later discussed
in \cite{Mueller} that interference effects in bosonic systems are particularly
strong due to the constructive interference
between all low-energy single- and many-boson cotunnelling processes, which leads,
e.g., to a huge positive magnetoresistance\cite{Shahar:giantMR, Kapitulnik:giantMR}
and broadens the superfluid phase of the bosons.

dc Transport at strong disorder requires inelastic hopping of charge carriers through the whole sample,
assisted by absorption/emission of neutral excitations, e.g., phonons\cite{Mott:book,Shklovskii:book}.
ac Conductivity may come from local excitations and
does not vanish even in the limit of infinitesimal dissipation\cite{Efros:acreview}.
The simplest local excitation is a particle-antiparticle dipole created by external field
on a pair of neighbouring sites\cite{Holstein:fewsites}-- impurities or quasi-localised states.
To study ac magnetoresistance and the Hall effect in hopping insulators\cite{Imry:Hall, Imry:Hallins, Viehweger:magnetocond, Drichko:twosite} 
one has to consider processes on three or more sites\cite{Holstein:fewsites}.
So far studies of magnetotransport and interference phenomena concerned fermionic
charge carriers (electrons) and did not address the influence of quantum statistics.

In this paper we calculate the ac conductivity of strongly disordered fermionic
and bosonic insulators at zero temperature. We demonstrate that coherent processes on three or more
sites cause the interference correction to the conductivity, that depends on the
statistics of the charge carriers. For bosons the correction
is always positive, while for fermions it may be positive or negative if the conductivity is
dominated by particles or holes respectively. The latter holds even in the case of non-interacting fermions.
We calculate the conductivity explicitly for sufficiently large frequencies assuming
that the density of the charge carrier states is constant at respective energies.

Our results apply equally to charged particles in disordered media and to neutral cold atoms in optical
lattices. In the latter case the conductivity should be understood as a response to a tilt of the external
potential.

{\it Model.} Strongly disordered insulator may be modelled as an array of random-energy sites,
e.g., electron puddles, grains, impurities, quasi-localised states, with a weak intersite tunnelling.
We assume that the tunnelling conserves particle spin and that the characteristic on-site interactions
significantly exceed the characteristic energies of conducting excitations. Then
one may consider spinless particles, multiplying the conductivity by the spin degeneracy in the
end. The generic Hamiltonian of the charge carriers reads
\begin{equation}
\hat{\mathcal{H}}=\sum_{\mathbf{r}}U_{\mathbf{r}}(n_{\mathbf{r}})-\sum_{%
\mathbf{r}\neq \mathbf{r}^{\prime }}t_{\mathbf{r}\mathbf{r}^{\prime }}{\hat{b%
}}_{\mathbf{r}}^{\dagger }{\hat{b}}_{\mathbf{r}^{\prime }},  
\label{Haminit}
\end{equation}
where $\hat b_\br^\dagger$ and $\hat b_\br$ are the (bosonic or fermionic) particle creation
and annihilation operators on site $\br$, $t_{\br\br^\prime}$-- the intersite tunnelling element;
$U_\br(n_\br)$ is the energy
of $n_\br=\hat b_\br^\dagger\hat b_\br$ particles on site $\br$-- a large random
function of $n_\br$, which accounts for the interaction of particles and
a random potential.
In particular, these energies prevent superfluidity in the case of bosonic particles.
The disorder manifests itself in the fluctuations 
of the on-site energies,
$U_\br(n_\br)$,
intersite couplings $t_{\br\br^\prime}$, and random locations of the sites.

The intersite interactions are assumed to be negligible, which corresponds to a system of neutral cold
atoms\cite{Bloch:review}, a Josephson network with large self-capacitances\cite{FazioVZant}, and can be also justified
in systems with weak Coulomb interactions, provided all relevant energies sufficiently
exceed the Coulomb gap\cite{Shklovskii:book}. We assume that each occupation number $n_\br$ 
uniquely determines the many-particle state on site $\br$ because of sufficiently strong
on-site interactions, leading to large energy gaps between different orbital states of the
$n_\br$ particles.
Eq.~(\ref{Haminit}) is a generic Hamiltonian that describes, for instance,
the Anderson model or the disordered Bose-Hubbard model
in the case of fermions or bosons respectively.

The strongly insulating regime requires the tunnelling $t_{\br\br^\prime}$ to be small
compared to the fluctuations of the on-site energies
and to sufficiently quickly decay with distance.
Thus, the groundstate of the Hamiltonian (\ref{Haminit})
is close to the groundstate of the on-site energies, that corresponds to
the configuration of integers $n_\br^0$ minimising the first term in the right-hand side
of Eq.~(\ref{Haminit}) \cite{Syzranov:RG}.
At low temperatures and frequencies only the lowest-energy excitations are
important on each site\cite{Syzranov:RG}, corresponding to the occupation numbers $n_\br^0+1$
and $n_\br^0-1$, which will be referred to as particle and hole excitations on site $\br$ respectively,
both in the case of bosons and fermions.

In the case of fermions the important lowest-energy excitations correspond to only one orbital state on each site,
the interaction
with the other electrons being equivalent to a static potential, and, without loss of generality,
we may assume that the occupation numbers take only values
$n_\br=0$ and $n_\br=1$. In the case of bosons we assume 
$n_\br^0\gg1$ for simplicity; however, the conclusions of this paper hold at arbitrary average
characteristic occupation numbers.

We calculate the conductivity assuming that the excitation density of states $\nu$ per site
is constant up to the characteristic cutoffs $E^p$ and $E^h$ of the particle and hole
excitation energies, and that the frequency exceeds the characteristic
matrix elements of the intersite couplings but is smaller than the characteristic excitation energies,
\begin{equation}
	J_{\br\br^\prime}\ll\omega\ll\nu^{-1}, max(E^p, E^h),
	\label{freqrange}
\end{equation}
where $J_{\br\br^\prime}=t_{\br\br^\prime}$ for fermionic charge carriers, while for
bosons $J_{\br\br^\prime}=t_{\br\br^\prime}(n_\br^0 n_{\br^\prime}^0)^{1/2}$.

A uniform external field $\bE(\omega)$ induces a current, $I_{12}(\omega)$, on each pair of sites $1$ and $2$,
which at low temperatures can be expressed through the retarded correlator of $I_{12}$ and currents
$I_{\br\br^\prime}$
on all the other pairs of sites using the Kubo formula:
\begin{equation}
	I_{12}(\omega)=i{(2\omega)}^{-1}\sum_{\br,\br^\prime}
	\Pi_{12,\br\br^\prime}(\omega)
	U_{\br\br^\prime} 
	\label{Kubo}
\end{equation}
\begin{equation}
	\Pi_{12,\br\br^\prime}(\omega)=
	- i \, \int_{0}^{\infty}
	\left\langle
	[ \hat I_{12}(t), \hat I_{\br\br^\prime}(0)]
	\right\rangle e^{i\omega t} {\rm d} \, t \, .
	\label{Pi}
\end{equation}
Here $\hat I_{\br\br^\prime}=iq\left(t_{\br^\prime\br}\hb_{\br^\prime}^\dagger\hb_\br
-t_{\br\br^\prime}\hb_{\br}^\dagger\hb_{\br^\prime}\right)$ and $U_{\br\br^\prime}=(\br^\prime-\br)\bE$
are the current operator and the effective voltage drop between
sites $\br$ ad $\br^\prime$, $q$-- the particle charge,
$\bE$-- the uniform amplitude
of the electric field.
In absence of external magnetic field the average current density 
evaluates
\begin{equation}
	j=\frac{1}{2}n^2\int\langle I_{\br\br^\prime}U_{\br\br^\prime}\rangle_{dis} |\bE|^{-1}
	d^d(\br^\prime-\br),
	\label{idensity}
\end{equation}
$n$ being the concentration of the sites, $\langle\ldots\rangle_{dis}$ is our convention for the disorder
averaging.


{\it Two-site conduction.} 
At infinitesimal dissipation the conduction comes from the resonant absorption of the electromagnetic
field quanta $\omega$ by local particle-antiparticle excitations on sparse pairs of sites\cite{Holstein:fewsites}.
Creating more complicated excitations is suppressed by the smallness of the tunnelling,
\begin{equation}
	\alpha\equiv\langle\nu \sum_{\br^\prime} J_{\br\br^\prime}\rangle_{dis}
	=n\nu\int \langle J_{\br\br^\prime}\rangle_{dis}d\br^\prime \ll1.
	\label{alpha}
\end{equation}
Clearly, the particle hopping between two sites is not affected by quantum statistics, except maybe
for the value of the coefficient before the hopping rate. We evaluate first the contribution of these
trivial two-site processes to the conductivity before analysing the interference and many-body corrections
to it. We consider the limit of low temperatures $T\ll\omega$.

%

Two-site processes correspond to the terms with $\br=1(2)$, $\br^\prime=2(1)$ in Eq.~(\ref{Kubo}).
Evaluating the correlator of the current
$I_{12}$ with itself we obtain the real part of the two-site conductance:
\begin{equation}
	G_{12}={\pi q^2}\omega^{-1}|J_{12}|^2\delta(E_2-E_1-\omega),
	\label{G12}
\end{equation}
where $E_1$ and $E_2$ are the energies of the particle on the respective sites, the ground state being close
to the particle residing on site $1$.
The delta-function in Eq.~(\ref{G12}) reflects the resonant absorption. Actually, in presence of a finite weak
dissipation it should be ascribed a certain width $\gamma$, the concentration of resonant pairs of sites
$\sim n\gamma\nu$ being very small.
Unless quenched disorder is long-correlated,
the intersite couplings $J_{12}=J_{|\br_2-\br_1|}$ and the on-site energies $E_{1,2}$ fluctuate independently.

The conductances (\ref{G12}) between pairs of sites lead to the conductivity
\begin{equation}
	\sigma_0={\pi n^2q^2}({2\omega d})^{-1}\int
	\langle J_\xi^2\rangle_{dis}\xi^2 \nu_{dip}(\omega,\xi) d^d\bxi,
\end{equation}
$\nu_{dip}(\omega,\xi)=2\langle\delta(E_2-E_1-\omega)\rangle_{dis}$ being the density of states
of dipole excitations, particle-hole pairs
of size $\xi$, factor $2$ accounting for the 2 possible
polarisations of a dipole at the same $E_1$ and $E_2$.

For a constant density of states $\nu$ of the single-site particle and hole excitations,
considered in this paper, we find $\nu_{dip}=2\nu^2\omega$, which yields a frequency-independent
conductivity
\begin{equation}
	\sigma_0=\pi n^2\nu^2q^2d^{-1}\int\langle J_\xi^2\rangle_{dis}\xi^2d^d\bxi.
	\label{cond0}
\end{equation}

Let us emphasise that the conductivity $\sigma_0(\omega)$ is constant
only in the frequency range under consideration.
At higher frequencies, violating the second of inequalities (\ref{freqrange}),
the conductivity decays due to the lack of sufficiently
high-energy states.
At smaller frequencies, when the first of inequalities (\ref{freqrange}) no longer holds, the conductivity
is described by the famous Mott's formula\cite{Mott:firstac, Efros:acreview} and
decreases $\propto\omega^2$.

Frequency-independent conductivity, Eq.~(\ref{cond0}), can be understood from the Mott's formula as follows.
A small voltage $V$, applied to a resonant pair of sites, $E_1-E_2=\omega$, with a small intersite coupling $J\ll\omega$,
makes a perturbation with the off-diagonal entry $\sim qVJ/(E_1-E_2)\propto\omega^{-1}$. The Mott's
formula, on the opposite, applies when the typical coupling exceeds the frequency, $\omega\lesssim\langle J\rangle_{dis}$.
Then the matrix element of the intersite transitions is non-perturbative in $J/\omega$ 
and does not depend on frequency (except, maybe, for a logarithmic factor).
Because the conductivity is quadratic in the matrix element,
$\sigma(\omega)\propto\sigma_{Mott}(\omega)\omega^{-2}=const$
at high frequencies under consideration, $\omega\gg\langle J\rangle_{dis}$.

{\it Three close sites} is the smallest cluster of sites that accounts for the interference
and non-trivial quantum statistics effects.
The contribution of larger clusters to the conductivity is suppressed due to the smallness
of the intersite tunnelling, Eq.~(\ref{alpha}).

There are two possibilities for the lowest-energy single-site excitations in a cluster of three sites:
1) a hole can occur on one site and an extra particle-- on each of the other two, or, vice versa,
2) a particle excitation can
occur on one site and a hole-- on each of the other two.

In the first case the dynamics of the three sites is equivalent to the hopping of
a single particle between these sites, described, both in the case of bosons and fermions,
by the effective Hamiltonian
\begin{equation} 
	H_{3}^{particle}=
	\left(
	\begin{array}{ccc}
		E_1 & -J_{12} & -J_{13} \\
		-J_{21} & E_2 & -J_{23} \\
		-J_{31} & -J_{32} & E_3
	\end{array}
	\right),
	\label{spartHam}
\end{equation}
where $E_\br$ is the energy of the particle on site $\br$, $E_1<E_2, E_3$.
The three-site Hamiltonian in the form (\ref{spartHam}) has been used in a number of works
(cf. e.g. \cite{Imry:Hall, Imry:Hallins, Entin:Hall}) to study the
Hall effect in hopping insulators.

{\it Single-hole hopping.} In the second case the dynamics of the three sites is equivalent
to the hopping of a lack of a particle between these sites. However, the effective Hamiltonian
will depend dramatically on the statistics of the charge carriers.

Indeed, the state of several fermions on several sites is antisymmetric with respect to the permutations of the fermions,
which makes the sign of the tunnelling element between two sites depend on the occupation numbers of the other sites,
$\langle 1_i0_k|\hb_i^\dagger\hb_k|0_i1_k\rangle=(-1)^{\sum_{j=i+1}^{k-1}n_j}$ for $i\leq k-2$ \cite{Landafshitz}.
For bosons all the signs are the same.
From Eq.~(\ref{Haminit}) we find the effective Hamiltonian which describes the hopping of a hole
on three sites:
\begin{equation} 
	H_{3}^{hole}=
	\left(
	\begin{array}{ccc}
		E_1 & -J_{21} & \mp J_{31} \\
		-J_{12} & E_2 & -J_{32} \\
		\mp J_{13} & -J_{23} & E_3
	\end{array}
	\right),
	\label{holeHam}
\end{equation}
where the upper and lower signs apply to bosons and fermions respectively,
$E_\br$ is the energy of a state with a hole on site $\br$, $E_1<E_2, E_3$.

The hopping of a single particle or hole is not a many-body problem effectively. 
Nevertheless, Eqs.~(\ref{spartHam}) and (\ref{holeHam}) show that the parameters
of the respective single-body Hamitonian {\it qualitatively depend on the statistics}:
for bosons (particles and holes) and fermionic particles the signs
of all tunnelling elements are the same, while for a fermionic hole the sign is alternating.
Below we demonstrate that this difference manifests itself,
under certain conditions, in the sign of the
interference correction to the ac conductivity.

{\it The currents between three sites} may be found, using Eqs.~(\ref{Kubo}) and (\ref{Pi}),
straightforwardly from the Hamiltonians (\ref{spartHam})
and (\ref{holeHam}). Because the interference correction
to the conductivity has a relative smallness
$\alpha$, Eq.~(\ref{alpha}), compared to the two-site contribution (\ref{cond0}),
it is sufficient to find the currents up to the third order in the small couplings $J$,
corresponding to the first-order-in-$\alpha$ correction to the two-site current
$I_{12}=G_{12}U_{12}$, Eq.~(\ref{G12}).

Again, conduction requires some resonant excitation be present on the three sites.
Below we assume that sites $1$ and $2$ in the cluster under consideration
are resonant, i.e. $E_2-E_1\approx\omega$. The probability of finding one more
resonant pair in the same cluster $\sim\gamma\nu$ is negligible.

In absence of external magnetic field all the couplings
may be chosen real and positive, $J_{\br\br^\prime}=|J_{\br\br^\prime}|$. Diagonalising
the Hamiltonians (\ref{spartHam}) and (\ref{holeHam}), and evaluating the
correlators of the currents, from Eqs.~(\ref{Kubo}) and (\ref{Pi}) we find  
the currents due to the hopping of particles and holes on three sites: 
\begin{eqnarray}
	I_{12}={\pi q^2}{\omega}^{-1}\delta(E_2-E_1-\omega)
	\Bigg\{
	|J_{12}|^2U_{12}
	\nonumber \\
	\pm J_{12}J_{23}J_{13}
	\left[\frac{U_{13}}{E_3-E_2}+\frac{U_{32}}{E_3-E_1}\right]
	\Bigg\},
	\label{I12}
\end{eqnarray}
\begin{eqnarray}
	I_{13}=\pm{\pi q^2}{\omega}^{-1}\delta(E_2-E_1-\omega)
	J_{12}J_{23}J_{13}\frac{U_{12}}{E_3-E_2},
	\label{I13}
	\\
	I_{23}=\pm{\pi q^2}{\omega}^{-1}\delta(E_2-E_1-\omega)
	J_{12}J_{23}J_{13}\frac{U_{12}}{E_3-E_1},
	\label{I23}
\end{eqnarray}
where the upper signs apply to fermionic particles and to bosonic particles and holes,
the lower signs- to fermionic holes only.

The first line of Eq.~(\ref{I12}) is the current due to the processes on the two
resonant sites, while the second line
of Eq.~(\ref{I12}) and Eqs. (\ref{I13}) and (\ref{I23}) are respectively the interference correction
to the two-site current and the extra currents due to the presence of the third site. 

{\it Bosonic interference correction.} In most three-site clusters under consideration,
the excitation energy $E_3$ on the third site
greatly exceeds the energies $E_1, E_2\sim\omega$
on the two resonant sites due to the condition (\ref{freqrange}).

In the case of bosons the direct current between sites $1$ and $2$ is increased by the presence of the
third site at $E_3\gg E_1, E_2$, yet the extra current $I_{13}\approx I_{32}$, flowing through
the third site, additionally enhances transport between sites $1$ and $2$. This reflects the
general principle that all low-energy cotunnelling processes in a bosonic system interfere constructively,
effectively enhancing low-energy transport\cite{Spivak:statistics, Syzranov:RG}.

To find the conductivity 
it is convenient to average first the currents in a three-site cluster with respect to
the energy of the excitation on site $3$. The positions of sites fluctuate independently
of the couplings and can be averaged separately. Using Eqs.~(\ref{idensity}), (\ref{I12})-(\ref{I23})
we find
the interference correction to the conductivity $\sigma_0$, Eq.~(\ref{cond0}),
for a bosonic system
with a constant density of states of single-site excitations at $\omega\ll E^{p,h}$:
\begin{eqnarray}
	\sigma_1(\omega)=
	A\log(E^p/\omega)+A\log(E^h/\omega)
	\label{sigma1boson}
	\\
	A=2\pi d^{-1}n^3\nu^3 
	\int \left\langle J_\zeta J_\xi
	J_{|\bxi-\bzeta|}\right\rangle_{dis} \xi^2d^d\bxi d^d\bzeta.
	\label{A}
\end{eqnarray}
The logarithms in Eq.~(\ref{sigma1boson}) come from integrating
Eqs.~(\ref{I12})-(\ref{I23}) wrt the third-site energies
at $\omega\ll E_3\ll E^{p,h}$ and must be smaller than $\alpha^{-1}$
to ensure the validity of the perturbation theory. The frequency $\omega$
serves as an effective low-energy cutoff due to the saturation of the average current corrections
at small energies on the third site, $0<E_3-E_1\lesssim E_2-E_1=\omega$.
If the frequency $\omega$ exceeds the energy cutoffs $E^p$ or $E^h$ of the particle
or hole excitations, the respective contribution in Eq.~(\ref{sigma1boson}) vanishes. 

Eqs.~(\ref{sigma1boson}) and (\ref{A}) show that in a bosonic system both particle and hole
contributions to the interference correction to the conductivity are positive.

{\it Fermionic interference correction.} The single-particle and single-hole processes contribute
to the interference correction with different signs in the case of fermions:
\begin{equation}
	\sigma_1=A \log[\max(E^p,\omega)/\omega]-A \log[\max(E^h,\omega)/\omega],
	\label{sigma1fermion}
\end{equation}
where constant $A$ is defined by Eq.~(\ref{A}). 
The cutoffs $E^p$ and $E^h$
depend on the level of doping of the insulator.
If the band of the localised states is nearly empty, then $E^p\gg E^h$,
and the interference correction to the conductivity is positive. An almost
filled band corresponds to $E^p\ll E^h$, leading to a negative correction
(\ref{sigma1fermion}).

In the whole frequency range, Eq.~(\ref{freqrange}), the correction is positive or negative
if the conduction is dominated respectively by particles or holes.
At sufficiently small frequencies, $\omega\ll E^{p,h}$, the correction is frequency-independent,
$\sigma_1=A\log(E^p/E^h)$. Despite this, it can be separated from the trivial two-site
contribution $\sigma_0$ using the suppression of interference by magnetic field,
as we discuss below.
If the frequency lies between the two cutoffs,
$E^{p,h}\ll\omega\ll E^{h,p}$, the conduction 
is dominated by the excitations with the higher cutoff, $\sigma_1=\pm A\log(E^{p,h}/\omega)$.
Let us notice, that in the latter regime of an effectively one type of fermionic charge carriers,
the frequency dependency of $\sigma_1$ is the same as for bosons, Eq.~(\ref{sigma1boson}),
but has a different sign if the charge carriers are holes.

{\it Suppression by gauge fields.} In presence of magnetic field the tunnelling elements of charged particles
acquire phases
\begin{equation}
	J_{\br\br^\prime}=|J_{\br\br^\prime}|\exp\left(iqc^{-1}\int_{\br}^{\br^\prime}\bA(\bxi)d\bxi\right),
\end{equation}
where $\bA$ is the vector potential, related to the magnetic field, and we neglected the modification
of the on-site wavefunctions. In the case of neutral atoms in optical lattices, the role of magnetic field
may be played by artificial gauge fields, induced by rotation or the Berry phases of atomic levels\cite{Bloch:review}.

The phase factors fluctuate randomly, due to the random relative positions
of the sites. This does not affect the two-site contribution to the conductivity $\sigma_0$ but
destroys the interference corrections to it if the magnetic flux through a characteristic
three-site cluster exceeds one flux quantum\cite{Syzranov:RG}, $qB\lambda^2\gg1$,
where $\lambda$ is the characteristic radius of the coupling $J_{\br\br^\prime}$.

Charge carriers with spin projections $s_z$ acquire additional energies $-s_z B$ in magnetic
field, which has a negligible effect on the conductivity provided $|s_z B|\ll E^{p,h}$.
Thus, the interference correction determines the sign of the magnetoresistance: positive for bosons,
and positive or negative for fermions depending on the doping level.

{\it Discussion.} We studied ac transport in strongly disordered systems.
The conductivity is dominated by statistics-independent processes on pairs of sites, while 
larger site clusters give rise to the interference
correction, which is always positive for bosons and can have
either sign for fermions.

Indeed, different commutation rules for bosonic
and fermionic variables lead to different character of interference effects\cite{Spivak:statistics}.
All low-energy bosonic processes interfere constructively,
effectively enhancing transport\cite{Syzranov:RG} and leading to a positive bosonic interference correction.

Fermionic many-body processes can have different signs.
We have shown that particle-dominated fermionic processes give a positive correction, hole-dominated-- negative, even
in the case of non-interacting fermions.
At first glance, transport of non-interacting particles and holes is a single-body problem and should not depend on the
statistics of the charge carriers. Nevertheless, the parameters of the effective single-body Hamiltonians
depend on the statistics, which in the case of fermions manifests itself in the alternating signs
of the effective intersite couplings, Eq.~(\ref{holeHam}).

We have calculated the conductivity explicitly,
Eqs.~(\ref{cond0}), (\ref{sigma1boson}) and (\ref{sigma1fermion}), under the assumptions
of a constant density of states, low temperature, and negligible intersite interactions at relevant energies.
However, our qualitative conclusions for
the sign of the ac magnetoresistance are valid for arbitrary interactions, densities of states,
and temperatures smaller than the on-site energy fluctuations. 
Experimentally one can suppress the interference correction by magnetic field or artificial gauge fields
and thus verify its sign.
In a strong fermionic insulator the interference contribution to the ac magnetoresistance
changes sign when changing the doping level.

Our results can be tested straightforwardly in experiments on 
spin-polarised fermionic\cite{Kondov:disordered} and bosonic\cite{Jendr:disordered} cold atoms
localised in incommensurate optical lattices\cite{Roati:incommensurate} or in a random
potential\cite{Kondov:disordered, Jendr:disordered}.
The interactions in these systems are short-ranged, yet the disorder and the interaction strength
are easily controlled\cite{Bloch:review}. These systems may be in the insulating state, which implies the smallness
of the intersite coupling compared to the fluctuations of the on-site energies.
The respective frequencies, cf. Eq.~(\ref{freqrange}), lie in an easily accessible kilohertz range
and exceed the typical temperatures $T\leq10nK$. Thus, all the assumptions of this paper are fulfilled in
such experiments, which allows one to verify not only our qualitative conclusions for the signs
of magnetoresistance but also the explicit dependencies of the conductivity on frequency.

The discussed effects can be observed also in disordered superconductive
films in the insulating state. Again, strong insulation indicates of the
smallness of the tunnelling. The characteristic frequencies used in the recent experiments, Refs.~\cite{Crane:ac1, Crane:ac2},
$\omega\sim10-100GHz$, significantly exceed achievable temperatures $T\sim.1K\sim1GHz$.
The behaviour of the ac conductivity
in magnetic field would
help one to identify which charge carriers, electrons or Cooper pairs, dominate transport in those materials.

{\it Acknowledgements.} 
We appreciate insightful discussions with B.L.~Altshuler, D.M.~Basko, and V.E.~Kravtsov.
We are particularly indebted to Yu.M.~Galperin for valuable discussions and suggestions on improving
the manuscript. This work has been financially supported by SFB Transregio 12.



\begin{thebibliography}{27}
\expandafter\ifx\csname natexlab\endcsname\relax\def\natexlab#1{#1}\fi
\expandafter\ifx\csname bibnamefont\endcsname\relax
  \def\bibnamefont#1{#1}\fi
\expandafter\ifx\csname bibfnamefont\endcsname\relax
  \def\bibfnamefont#1{#1}\fi
\expandafter\ifx\csname citenamefont\endcsname\relax
  \def\citenamefont#1{#1}\fi
\expandafter\ifx\csname url\endcsname\relax
  \def\url#1{\texttt{#1}}\fi
\expandafter\ifx\csname urlprefix\endcsname\relax\def\urlprefix{URL }\fi
\providecommand{\bibinfo}[2]{#2}
\providecommand{\eprint}[2][]{\url{#2}}

\bibitem[{\citenamefont{Shklovskii and Spivak}(1990)}]{ShklovskiiSpivak:review}
\bibinfo{author}{\bibfnamefont{B.~L.} \bibnamefont{Shklovskii}}
  \bibnamefont{and} \bibinfo{author}{\bibfnamefont{B.~Z.}
  \bibnamefont{Spivak}}, in \emph{\bibinfo{booktitle}{Hopping transport in
  solids}}, edited by \bibinfo{editor}{\bibfnamefont{M.}~\bibnamefont{Pollak}}
  \bibnamefont{and} \bibinfo{editor}{\bibfnamefont{B.~L.}
  \bibnamefont{Shklovskii}} (\bibinfo{publisher}{North-Holland, Amsterdam},
  \bibinfo{year}{1990}).

\bibitem[{\citenamefont{Bloch et~al.}(2008)\citenamefont{Bloch, Dalibard, and
  Zwerger}}]{Bloch:review}
\bibinfo{author}{\bibfnamefont{I.}~\bibnamefont{Bloch}},
  \bibinfo{author}{\bibfnamefont{J.}~\bibnamefont{Dalibard}}, \bibnamefont{and}
  \bibinfo{author}{\bibfnamefont{W.}~\bibnamefont{Zwerger}},
  \bibinfo{journal}{Rev. Mod. Phys.} \textbf{\bibinfo{volume}{80}},
  \bibinfo{pages}{885} (\bibinfo{year}{2008}).

\bibitem[{\citenamefont{Sambandamurthy
  et~al.}(2005)\citenamefont{Sambandamurthy, Engel, Johansson, Peled, and
  Shahar}}]{Shahar:superinsulator}
\bibinfo{author}{\bibfnamefont{G.}~\bibnamefont{Sambandamurthy}},
  \bibinfo{author}{\bibfnamefont{L.~W.} \bibnamefont{Engel}},
  \bibinfo{author}{\bibfnamefont{A.}~\bibnamefont{Johansson}},
  \bibinfo{author}{\bibfnamefont{E.}~\bibnamefont{Peled}}, \bibnamefont{and}
  \bibinfo{author}{\bibfnamefont{D.}~\bibnamefont{Shahar}},
  \bibinfo{journal}{Phys. Rev. Lett.} \textbf{\bibinfo{volume}{94}},
  \bibinfo{pages}{017003} (\bibinfo{year}{2005}).

\bibitem[{\citenamefont{Baturina et~al.}(2007)\citenamefont{Baturina, Mironov,
  Vinokur, Baklanov, and Strunk}}]{Baturina:superinsulator}
\bibinfo{author}{\bibfnamefont{T.~I.} \bibnamefont{Baturina}},
  \bibinfo{author}{\bibfnamefont{A.~Y.} \bibnamefont{Mironov}},
  \bibinfo{author}{\bibfnamefont{V.~M.} \bibnamefont{Vinokur}},
  \bibinfo{author}{\bibfnamefont{M.~R.} \bibnamefont{Baklanov}},
  \bibnamefont{and} \bibinfo{author}{\bibfnamefont{C.}~\bibnamefont{Strunk}},
  \bibinfo{journal}{Phys. Rev. Lett.} \textbf{\bibinfo{volume}{99}},
  \bibinfo{pages}{257003} (\bibinfo{year}{2007}).

\bibitem[{\citenamefont{Gantmakher and
  Dolgopolov}(2010)}]{Gantmakher:SITreview}
\bibinfo{author}{\bibfnamefont{V.~F.} \bibnamefont{Gantmakher}}
  \bibnamefont{and} \bibinfo{author}{\bibfnamefont{V.~T.}
  \bibnamefont{Dolgopolov}}, \bibinfo{journal}{Usp. Fiz. Nauk}
  \textbf{\bibinfo{volume}{53}}, \bibinfo{pages}{1} (\bibinfo{year}{2010}).

\bibitem[{\citenamefont{Syzranov et~al.}(2012)\citenamefont{Syzranov, Moor, and
  Efetov}}]{Syzranov:RG}
\bibinfo{author}{\bibfnamefont{S.~V.} \bibnamefont{Syzranov}},
  \bibinfo{author}{\bibfnamefont{A.}~\bibnamefont{Moor}}, \bibnamefont{and}
  \bibinfo{author}{\bibfnamefont{K.~B.} \bibnamefont{Efetov}},
  \bibinfo{journal}{Phys. Rev. Lett.} \textbf{\bibinfo{volume}{108}},
  \bibinfo{pages}{256601} (\bibinfo{year}{2012}).

\bibitem[{\citenamefont{Mueller}(2012)}]{Mueller}
\bibinfo{author}{\bibfnamefont{M.}~\bibnamefont{Mueller}}
  (\bibinfo{year}{2012}), \bibinfo{note}{arXiv:1109.0245v2}.

\bibitem[{\citenamefont{Sambandamurthy
  et~al.}(2004)\citenamefont{Sambandamurthy, Engel, Johansson, and
  Shahar}}]{Shahar:giantMR}
\bibinfo{author}{\bibfnamefont{G.}~\bibnamefont{Sambandamurthy}},
  \bibinfo{author}{\bibfnamefont{L.~W.} \bibnamefont{Engel}},
  \bibinfo{author}{\bibfnamefont{A.}~\bibnamefont{Johansson}},
  \bibnamefont{and} \bibinfo{author}{\bibfnamefont{D.}~\bibnamefont{Shahar}},
  \bibinfo{journal}{Phys. Rev. Lett.} \textbf{\bibinfo{volume}{92}},
  \bibinfo{pages}{107005} (\bibinfo{year}{2004}).

\bibitem[{\citenamefont{Steiner and Kapitulnik}(2005)}]{Kapitulnik:giantMR}
\bibinfo{author}{\bibfnamefont{M.}~\bibnamefont{Steiner}} \bibnamefont{and}
  \bibinfo{author}{\bibfnamefont{A.}~\bibnamefont{Kapitulnik}},
  \bibinfo{journal}{Physica C} \textbf{\bibinfo{volume}{422}},
  \bibinfo{pages}{16} (\bibinfo{year}{2005}).

\bibitem[{\citenamefont{Mott and Davis}(1979)}]{Mott:book}
\bibinfo{author}{\bibfnamefont{N.~F.} \bibnamefont{Mott}} \bibnamefont{and}
  \bibinfo{author}{\bibfnamefont{E.~A.} \bibnamefont{Davis}},
  \emph{\bibinfo{title}{Electronic processes in non-crystalline materials}}
  (\bibinfo{publisher}{Clarendon, Oxford}, \bibinfo{year}{1979}).

\bibitem[{\citenamefont{Shklovskii and L.Efros}(1984)}]{Shklovskii:book}
\bibinfo{author}{\bibfnamefont{B.~I.} \bibnamefont{Shklovskii}}
  \bibnamefont{and} \bibinfo{author}{\bibfnamefont{A.}~\bibnamefont{L.Efros}},
  \emph{\bibinfo{title}{Electronic properties of doped semiconductors}}
  (\bibinfo{publisher}{Springer, Heidelberg}, \bibinfo{year}{1984}).

\bibitem[{\citenamefont{Efros and Shklovskii}(1985)}]{Efros:acreview}
\bibinfo{author}{\bibfnamefont{A.~L.} \bibnamefont{Efros}} \bibnamefont{and}
  \bibinfo{author}{\bibfnamefont{B.~L.} \bibnamefont{Shklovskii}}, in
  \emph{\bibinfo{booktitle}{Electron-electron interactions in disordered
  systems}}, edited by \bibinfo{editor}{\bibfnamefont{A.~L.}
  \bibnamefont{Efros}} \bibnamefont{and}
  \bibinfo{editor}{\bibfnamefont{M.}~\bibnamefont{Pollak}}
  (\bibinfo{publisher}{North-Holland, Amsterdam}, \bibinfo{year}{1985}).

\bibitem[{\citenamefont{Holstein}(1961)}]{Holstein:fewsites}
\bibinfo{author}{\bibfnamefont{T.}~\bibnamefont{Holstein}},
  \bibinfo{journal}{Phys. Rev.} \textbf{\bibinfo{volume}{124}},
  \bibinfo{pages}{1329} (\bibinfo{year}{1961}).

\bibitem[{\citenamefont{Imry}(1993)}]{Imry:Hall}
\bibinfo{author}{\bibfnamefont{Y.}~\bibnamefont{Imry}}, \bibinfo{journal}{Phys.
  Rev. Lett.} \textbf{\bibinfo{volume}{71}}, \bibinfo{pages}{1868}
  (\bibinfo{year}{1993}).

\bibitem[{\citenamefont{Entin-Wohlman et~al.}(1995)\citenamefont{Entin-Wohlman,
  Aronov, Levinson, and Imry}}]{Imry:Hallins}
\bibinfo{author}{\bibfnamefont{O.}~\bibnamefont{Entin-Wohlman}},
  \bibinfo{author}{\bibfnamefont{A.~G.} \bibnamefont{Aronov}},
  \bibinfo{author}{\bibfnamefont{Y.}~\bibnamefont{Levinson}}, \bibnamefont{and}
  \bibinfo{author}{\bibfnamefont{Y.}~\bibnamefont{Imry}},
  \bibinfo{journal}{Phys. Rev. Lett.} \textbf{\bibinfo{volume}{75}},
  \bibinfo{pages}{4094} (\bibinfo{year}{1995}).

\bibitem[{\citenamefont{Viehweger and Efetov}(1991)}]{Viehweger:magnetocond}
\bibinfo{author}{\bibfnamefont{O.}~\bibnamefont{Viehweger}} \bibnamefont{and}
  \bibinfo{author}{\bibfnamefont{K.~B.} \bibnamefont{Efetov}},
  \bibinfo{journal}{Phys. Rev. B} \textbf{\bibinfo{volume}{44}},
  \bibinfo{pages}{1168} (\bibinfo{year}{1991}).

\bibitem[{\citenamefont{Drichko et~al.}(2000)\citenamefont{Drichko, Diakonov,
  Smirnov, Galperin, and Toropov}}]{Drichko:twosite}
\bibinfo{author}{\bibfnamefont{I.~L.} \bibnamefont{Drichko}},
  \bibinfo{author}{\bibfnamefont{A.~M.} \bibnamefont{Diakonov}},
  \bibinfo{author}{\bibfnamefont{I.~Y.} \bibnamefont{Smirnov}},
  \bibinfo{author}{\bibfnamefont{Y.~M.} \bibnamefont{Galperin}},
  \bibnamefont{and} \bibinfo{author}{\bibfnamefont{A.~I.}
  \bibnamefont{Toropov}}, \bibinfo{journal}{Phys. Rev. B}
  \textbf{\bibinfo{volume}{62}}, \bibinfo{pages}{7470} (\bibinfo{year}{2000}).

\bibitem[{\citenamefont{Fazio and van~der Zant}(2001)}]{FazioVZant}
\bibinfo{author}{\bibfnamefont{R.}~\bibnamefont{Fazio}} \bibnamefont{and}
  \bibinfo{author}{\bibfnamefont{H.}~\bibnamefont{van~der Zant}},
  \bibinfo{journal}{Phys. Rep.} \textbf{\bibinfo{volume}{355}},
  \bibinfo{pages}{235} (\bibinfo{year}{2001}).

\bibitem[{\citenamefont{Mott}(1970)}]{Mott:firstac}
\bibinfo{author}{\bibfnamefont{N.~F.} \bibnamefont{Mott}},
  \bibinfo{journal}{Phil. Mag.} \textbf{\bibinfo{volume}{22}},
  \bibinfo{pages}{7} (\bibinfo{year}{1970}).

\bibitem[{\citenamefont{Entin-Wohlman et~al.}(2005)\citenamefont{Entin-Wohlman,
  Aharony, Galperin, Kozub, and Vinokur}}]{Entin:Hall}
\bibinfo{author}{\bibfnamefont{O.}~\bibnamefont{Entin-Wohlman}},
  \bibinfo{author}{\bibfnamefont{A.}~\bibnamefont{Aharony}},
  \bibinfo{author}{\bibfnamefont{Y.~M.} \bibnamefont{Galperin}},
  \bibinfo{author}{\bibfnamefont{V.~I.} \bibnamefont{Kozub}}, \bibnamefont{and}
  \bibinfo{author}{\bibfnamefont{V.}~\bibnamefont{Vinokur}},
  \bibinfo{journal}{Phys. Rev. Lett.} \textbf{\bibinfo{volume}{95}},
  \bibinfo{pages}{086603} (\bibinfo{year}{2005}).

\bibitem[{\citenamefont{Landau and Lifshitz}(1977)}]{Landafshitz}
\bibinfo{author}{\bibfnamefont{L.~D.} \bibnamefont{Landau}} \bibnamefont{and}
  \bibinfo{author}{\bibfnamefont{E.~M.} \bibnamefont{Lifshitz}},
  \emph{\bibinfo{title}{Quantum mechanics non-relativistic theory (Course of
  theoretical physics, volume 3)}} (\bibinfo{publisher}{Pergamon, Oxford},
  \bibinfo{year}{1977}), \bibinfo{note}{\S{65}}.

\bibitem[{\citenamefont{Zhao et~al.}(1991)\citenamefont{Zhao, Spivak, Gelfand,
  and Feng}}]{Spivak:statistics}
\bibinfo{author}{\bibfnamefont{H.~L.} \bibnamefont{Zhao}},
  \bibinfo{author}{\bibfnamefont{B.~Z.} \bibnamefont{Spivak}},
  \bibinfo{author}{\bibfnamefont{M.~P.} \bibnamefont{Gelfand}},
  \bibnamefont{and} \bibinfo{author}{\bibfnamefont{S.}~\bibnamefont{Feng}},
  \bibinfo{journal}{Phys. Rev. B} \textbf{\bibinfo{volume}{44}},
  \bibinfo{pages}{10760} (\bibinfo{year}{1991}).

\bibitem[{\citenamefont{Kondov et~al.}(2011)\citenamefont{Kondov, McGehee,
  Zirbel, and DeMarco}}]{Kondov:disordered}
\bibinfo{author}{\bibfnamefont{S.~S.} \bibnamefont{Kondov}},
  \bibinfo{author}{\bibfnamefont{W.~R.} \bibnamefont{McGehee}},
  \bibinfo{author}{\bibfnamefont{J.~J.} \bibnamefont{Zirbel}},
  \bibnamefont{and} \bibinfo{author}{\bibfnamefont{B.}~\bibnamefont{DeMarco}},
  \bibinfo{journal}{Science} \textbf{\bibinfo{volume}{334}},
  \bibinfo{pages}{66} (\bibinfo{year}{2011}).

\bibitem[{\citenamefont{Jendrzejewski et~al.}(2012)\citenamefont{Jendrzejewski,
  Bernard, M{\"u}ller, Cheinet, Josse, Piraud, Pezze, Sanchez-Palencia, Aspect,
  and Bouyer}}]{Jendr:disordered}
\bibinfo{author}{\bibfnamefont{F.}~\bibnamefont{Jendrzejewski}},
  \bibinfo{author}{\bibfnamefont{A.}~\bibnamefont{Bernard}},
  \bibinfo{author}{\bibfnamefont{K.}~\bibnamefont{M{\"u}ller}},
  \bibinfo{author}{\bibfnamefont{P.}~\bibnamefont{Cheinet}},
  \bibinfo{author}{\bibfnamefont{V.}~\bibnamefont{Josse}},
  \bibinfo{author}{\bibfnamefont{M.}~\bibnamefont{Piraud}},
  \bibinfo{author}{\bibfnamefont{L.}~\bibnamefont{Pezze}},
  \bibinfo{author}{\bibfnamefont{L.}~\bibnamefont{Sanchez-Palencia}},
  \bibinfo{author}{\bibfnamefont{A.}~\bibnamefont{Aspect}}, \bibnamefont{and}
  \bibinfo{author}{\bibfnamefont{P.}~\bibnamefont{Bouyer}},
  \bibinfo{journal}{Nature Phys.} \textbf{\bibinfo{volume}{8}},
  \bibinfo{pages}{398} (\bibinfo{year}{2012}).

\bibitem[{\citenamefont{Roati et~al.}(2008)\citenamefont{Roati, {D'Errico},
  Fallani, Fattori, Fort, Zaccanti, Modugno, Modugno, and
  Inguscio}}]{Roati:incommensurate}
\bibinfo{author}{\bibfnamefont{G.}~\bibnamefont{Roati}},
  \bibinfo{author}{\bibfnamefont{C.}~\bibnamefont{{D'Errico}}},
  \bibinfo{author}{\bibfnamefont{L.}~\bibnamefont{Fallani}},
  \bibinfo{author}{\bibfnamefont{M.}~\bibnamefont{Fattori}},
  \bibinfo{author}{\bibfnamefont{C.}~\bibnamefont{Fort}},
  \bibinfo{author}{\bibfnamefont{M.}~\bibnamefont{Zaccanti}},
  \bibinfo{author}{\bibfnamefont{G.}~\bibnamefont{Modugno}},
  \bibinfo{author}{\bibfnamefont{M.}~\bibnamefont{Modugno}}, \bibnamefont{and}
  \bibinfo{author}{\bibfnamefont{M.}~\bibnamefont{Inguscio}},
  \bibinfo{journal}{Nature} \textbf{\bibinfo{volume}{453}},
  \bibinfo{pages}{895} (\bibinfo{year}{2008}).

\bibitem[{\citenamefont{Crane et~al.}(2007{\natexlab{a}})\citenamefont{Crane,
  Armitage, Johansson, Sambandamurthy, Shahar, and Gruner}}]{Crane:ac1}
\bibinfo{author}{\bibfnamefont{R.~W.} \bibnamefont{Crane}},
  \bibinfo{author}{\bibfnamefont{N.~P.} \bibnamefont{Armitage}},
  \bibinfo{author}{\bibfnamefont{A.}~\bibnamefont{Johansson}},
  \bibinfo{author}{\bibfnamefont{G.}~\bibnamefont{Sambandamurthy}},
  \bibinfo{author}{\bibfnamefont{D.}~\bibnamefont{Shahar}}, \bibnamefont{and}
  \bibinfo{author}{\bibfnamefont{G.}~\bibnamefont{Gruner}},
  \bibinfo{journal}{Phys. Rev. B} \textbf{\bibinfo{volume}{75}},
  \bibinfo{pages}{094506} (\bibinfo{year}{2007}{\natexlab{a}}).

\bibitem[{\citenamefont{Crane et~al.}(2007{\natexlab{b}})\citenamefont{Crane,
  Armitage, Johansson, Sambandamurthy, Shahar, and Gruner}}]{Crane:ac2}
\bibinfo{author}{\bibfnamefont{R.~W.} \bibnamefont{Crane}},
  \bibinfo{author}{\bibfnamefont{N.~P.} \bibnamefont{Armitage}},
  \bibinfo{author}{\bibfnamefont{A.}~\bibnamefont{Johansson}},
  \bibinfo{author}{\bibfnamefont{G.}~\bibnamefont{Sambandamurthy}},
  \bibinfo{author}{\bibfnamefont{D.}~\bibnamefont{Shahar}}, \bibnamefont{and}
  \bibinfo{author}{\bibfnamefont{G.}~\bibnamefont{Gruner}},
  \bibinfo{journal}{Phys. Rev. B} \textbf{\bibinfo{volume}{75}},
  \bibinfo{pages}{184530} (\bibinfo{year}{2007}{\natexlab{b}}).

\end{thebibliography}
\end{document}